\documentclass[12pt]{article}

0

\begin{document}
\newcommand {\be}{\begin{equation}}
\newcommand {\ee}{\end{equation}}
\newcommand {\bea}{\begin{array}}
\newcommand {\cl}{\centerline}
\newcommand {\eea}{\end{array}}
\def\simlt{\stackrel{<}{{}_\sim}}
\def\simgt{\stackrel{>}{{}_\sim}}
\def\IP{\relax{\rm I\kern-.18em P}}
\def\CI{case {\bf (I) }}
\def\CII{case {\bf (II) }}
\def\sugra{supergravity }

\renewcommand {\theequation}{\thesection.\arabic{equation}}
\renewcommand {\thefootnote}{\fnsymbol{footnote}}
\newcommand {\newsection}{\setcounter{equation}{0}\section}
\csname @addtoreset\endcsname{equation}{section}

\def\ba{\begin{eqnarray}}
\def\ea{\end{eqnarray} }
\def\o{\over} 
\def \a {\alpha }
\def\th{\theta }
\def\s{\sigma}
\def\t{\tau }
\def\p{\partial }
\def\del{\partial }

\baselineskip 0.65 cm

\begin{flushright}
IC/2000/139\\
hep-th/0009141
\vskip 14mm 
\end{flushright}

\begin{center}
{\LARGE{ Strong Coupling Effects in Non-commutative Spaces\\
 {}From OM Theory and Supergravity}}
\vskip 14mm

{\bf {\large{J. G. Russo$^{1}$ and M. M. Sheikh-Jabbari$^2$}}}

\vspace{12 mm}

$^1${\it Departamento de F\'\i sica, Universidad de Buenos Aires\\
Ciudad Universitaria, 1428 Buenos Aires}\\
{\tt russo@df.uba.ar  }\\

$^2${\it The Abdus Salam International Center for Theoretical Physics\\
Strada Costiera, 11. 34014, Trieste, Italy}\\
{\tt jabbari@ictp.trieste.it  }\\

\end{center}

\vskip 8mm
\begin{center}

{\bf Abstract}

\end{center}

We show that a four-parameter class of 3+1 dimensional NCOS
theories can be obtained by dimensional reduction on a general 2-torus
from OM theory. 
Compactifying two spatial directions of  NCOS theory 
on a 2-torus, we study the transformation properties under 
the $SO(2,2;Z)$ T-duality group. 
We then discuss non-perturbative configurations
of non-commutative super Yang-Mills theory.
In particular, we calculate the tension for magnetic monopoles and
(p,q) dyons
and exhibit their six-dimensional origin,
and construct a supergravity solution representing an instanton
in the gauge theory.  
We also compute the potential for a monopole-antimonopole 
in the supergravity approximation.

\phantom{\cite{CDS,SW,SST,GMMS,OM,BBSS}}

\phantom{\cite{br,gm,chen,H,km,luroy,RS,kawt,agm,bs,rv,ohta,har,LRS,AOR}}

\vspace{2 mm}

\newpage

\newsection{Introduction}
The observation that non-commutative gauge theories
can be obtained as limits of string theory in a background 
$B$ field \cite{CDS,SW} has led to a number of interesting results,
in particular, the emergence of
string theories without gravity 
near certain corners of the moduli space.
A simple example is the open string theory constructed
from a D3 brane in the presence of a near critical electric field
\cite{SST,GMMS}. It is essentially a free string theory with
an additional term in the propagator which gives rise to extra phase
factors in front of the scattering amplitudes, producing the structure of a
non-commutative open string theory (NCOS).

Recently, it was proposed that one can similarly construct
a six dimensional theory as the low energy theory of an M5 brane in the
presence of a near critical three-form field \cite{OM,BBSS},
which contains open membranes and is decoupled from gravity.
In view of the parallel picture in M-theory, it appears then natural to conjecture that all these string theories 
can be obtained
as different limits of this six-dimensional (OM) theory \cite{OM}.
Further discussions can be found in refs. \cite{br} -\cite{AOR}.

The simplest NCOS theory is specified by two parameters, 
namely the open string coupling $G_s$ and
$\a '$, where $\a '$ represents both the string and the non-commutative scale in the
$x_0$-$x_1$ plane.
In \cite{RS} the NCOS theory was generalized to a larger class of NCOS theories, which are classified by four
parameters: $\a' , \ G_s ,\  \theta ^{23}$ and $\chi $
(see \cite{OM,luroy,LRS} for related discussions). The parameter $\theta ^{23}$ is the non-commutative
scale in the $x_2$-$x_3$ plane, and $\chi $ --which originates from a nonzero value for the RR scalar field--
gives rise to a $\theta_{\rm QCD} $ term $\theta_{\rm QCD}\int F\tilde F $, $\theta_{\rm QCD}=\chi $,
in the low energy Yang-Mills theory. For $\chi $ irrational, it was found that
the $SL(2, Z )$ transformations of type IIB superstring theory map the NCOS theory into another one with different
parameters. However, when $\chi $ is rational, it was found that
there exists an $SL(2,Z)$ transformation which maps the NCOS theory to non-commutative SYM field theory. This
generalizes the proposal of \cite{GMMS} that NCOS theory with $ \theta^{23} =\chi =0$ is S-dual to NCSYM
theory. 

Compactifications of NCOS theories in their own turn lead to various T-duality groups. For the particular case of 
(3+1) dimensional NCOS compactified on a two torus, 
we show that the full $SO(2,2;Z)\simeq 
SL(2,Z)\times SL(2,Z)$ T-duality group connects  NCOS theories. 
This T-duality group combined with the S-duality $SL(2,Z)$ group yield
the usual U-duality group of closed strings on $T^2$; $SL(3,Z)\times SL(2,Z)$. 

Non-commutative gauge theories contain many interesting non-perturbative dynamical objects (see e.g. \cite{NS,Hashi,GN,gross}).
These field-theory configurations can be described in the brane picture e.g. in terms of D-strings
or D-instantons (or other branes) intersecting with the D3 brane. In particular, a D-string ending on a
D3-brane is a magnetic charge from the point of view of the gauge theory on the brane, located at the end of the
string. In this paper we will analyise some of these systems using both weak-coupling string theory and supergravity, thus
complementing the discussion of \cite{Hashi,GN}. 

This paper is organized as follows. In sect.~2 we show that
compactification of OM theory on a general 2-torus
leads to the 4-parameter class of NCOS theories mentioned above.
In section 3 we consider the toroidal compactification of NCOS theory and discuss  $SL(2,Z)\times SL(2,Z)$
T-duality transformations.
In sect.~4, we discuss supersymmetric solitons and their six-dimensional origin. 
In sect.~5 we compute 
the monopole-antimonopole potential in large $N$ non-commutative super Yang-Mills theory.

 
\newsection{Reduction of OM theory on a general torus}

In this section we shall consider a reduction
on a general torus along an arbitrary oblique direction.
This will lead to the general 4-parameter class of NCOS theories
in 3+1 dimensions
considered in \cite{RS}. In \cite{RS}, the reduction has been done
for a rectangular torus, giving 3+1 dimensional NCOS theories with $\a' , G_s,  \theta^{23} $, but $\theta_{\rm QCD}=\chi $ parameter equal to zero. Here we will complete the discussion by
considering dimensional reduction on an arbitrary torus, which will
permit to incorporate a non-zero  parameter $\chi $.

Thus we start with the M5-brane configuration of \cite{OM},
but we take a general (non-orthogonal) coordinate system in the 2-plane $x_4$-$x_5$, i.e.
\be
H_3 =H_{01\tilde 5}\ dx_0\wedge dx_1\wedge d\tilde x_5 +
H_{23\tilde 4}\ dx_2\wedge dx_3\wedge d\tilde x_4\ ,
\label{hhhtt}
\ee
$$
\tilde x_4=x_4 \cos\gamma + x_5 \sin\alpha\ ,\ \ \ 
\tilde x_5=- x_4 \sin\gamma +x_5 \cos\alpha\ ,
$$
where the coordinates $x_4$ and $x_5$ are periodic with periods
\be
x_5\equiv x_5+2\pi R_5 \ , \ \ \ \ \ \ x_4\equiv x_4+2\pi R_4^0\ .
\ee
The $x_4$ axes is rotated by an angle $\gamma$ with respect to $\tilde x_4$, and $x_5$ is the axes
which is rotated by $\alpha$ with respect to $\tilde x_5$. The axes $\tilde x_4,\ \tilde x_5$ are
orthogonal. The orthogonal rotation considered in \cite{RS} is recovered
by setting $\gamma=\alpha$.

The metric in tilde coordinates is 
$$
g_{\mu\nu}={\rm diag}(-\xi^2,\xi^2,\rho^2,\rho^2,\xi^2,\xi^2 )\ ,
$$
where we have rescaled some coordinates by constant factors $\xi,\rho $, which will be fixed later.
So, in the $x_4, \ x_5$ coordinates, the metric for the
$45$ plane is given by
\be
ds^2=\xi^2(dx_5^2+dx_4^2-2\sin(\gamma-\alpha)dx_4\ dx_5)\ ,
\ee 
and the $H$ field components are
\be
H_{015}=- \xi^3 \tanh\beta \cos\alpha \ ,
\ \ \ \ \ 
H_{014}=\xi^3 \tanh\beta \sin\gamma\ ,
\ee
\be
H_{234}= \rho^2\xi \sinh\beta \cos\gamma \ ,
\ \ \ \ \
H_{235}= \rho^2 \xi \sinh\beta \sin\alpha\ .
\ee

The theory should be decoupled from gravity in the limit $\beta\to \infty
$, with 
\be\bea{cc}
M_{\rm eff}^3={1\over 2} M_P ^3\bigg(\xi ^{3}+ H_{01\tilde 5}\bigg)=
{1\over 2} M_P^3 \xi^3 \bigg(1- \tanh\beta \bigg) ={\rm
fixed}\ , \\
M_P^3 \ \xi\rho^2 ={\rm fixed}\ .\ 
\eea\ee
Thus we set
\be
\xi=\xi_0 e^{2\beta/3} \ ,\ \ \ \ \rho =\rho_0 e^{-\beta/3}\ ,\ \ \ \ 
\ee
and scale $\sin\alpha$ so that
\be
\sinh\beta\ \sin\alpha=B={\rm fixed}\ . 
\ee
with fixed $\xi_0,\ \rho_0,\ B,\ $ and $M_P$. We see that
in the limit $\beta\to\infty $, 
the angle $\alpha \to 0$.

Next, we make dimensional reduction along $x_5$. This gives $N$ 
D4 branes with $B_{01}$, $B_{23}$ field components, and
a non-vanishing RR one-form component $A_\mu $ coming 
from the $45$ component of metric. The gauge field components are
\be\bea{cc}
B_{01}=M_PR_5 \ H_{015}=- M_PR_5\xi^3\tanh\beta\cos\a \ ,\\
B_{23}=M_PR_5\ H_{235} =M_PR_5 \rho^2 \xi\sinh\beta\sin\alpha\ ,\\
A_4=(M_PR_5)^{-1}\ \sin(\gamma-\alpha)\ .
\eea\ee
The other components $H_{014},\ H_{234}$ lead to non-zero components
 of the 3-form RR field given by
\be
A_{014}=H_{014}\ , \ \ \ \ \ 
A_{234}=H_{234}\ .
\ee
The 5d metric is
\be
g_{\mu\nu}^{\rm (A)}=g_s^{2/3}{\rm diag}(-\xi^2,\xi^2,\rho^2,\rho^2,\xi^2 \cos^2(\gamma-\alpha))\ ,
\ee
with 
\be\label{Meff}
g_s=(\xi R_5 M_P)^{3/2} =e^{\beta } (R_5 M_{\rm eff})^{3/2} \ ,\ 
\ \ \ \ \ 
l_s^2=\alpha'= {1\over M_{\rm eff}^3 R_5}\ .
\ee
Thus one gets the following electric and magnetic field components:
\be
E=B^0_{\ 1}=\tanh\beta\cos\a \ , \ \ \ \ \ B=B^2_{\ 3}= \sinh\beta\ \sin\alpha\ .
\ee
As $\beta\to\infty $ one gets $B^0_{\ 1}=1$ and 
$B^2_{\ 3}=B$. 
Let us write $x_4=R_4^0\theta_4$, with $\theta_4=\theta_4+2\pi $.
Then 
\be
g^{\rm (A)}_{44}=({R_4^0})^2\ M_P R_5 \xi^3 \cos^2(\gamma-\alpha)\ .
\ee
Now we perform T-duality in the $x_4$ direction, which gives $N$ D3 branes 
with non-vanishing NSNS and RR two-form components, and 
set
$$
\xi_0=[4M_PR_5 (1+B^2)]^{-1/3}\ ,\ \ \ \ \ \rho_0={2\xi_0}
\ .
$$
This gives the following closed string metric for the $3+1$ dimensional world-volume:
$$
g_{\mu\nu}={\rm diag}(-{1\over 1-E^2}\ ,{1\over 1-E^2}\ ,{1\over
1+B^2}\ ,{1\over 1+B^2})\ ,
$$
and an open string metric given by
$G_{\mu\nu}={\rm diag}(-1,1,1,1)\ $.
The type IIB closed string coupling constant, $g_s^{(B)}$ and RR scalar field, $\chi$ are
$$
g_s^{(B)}=g_s\ \sqrt{\alpha'\over g^{(A)}_{44}}={M_P R_5 \sqrt{\a'}\over R_4^0
\cos(\gamma-\alpha)}\ ,
$$
\be
\chi=A_4\ { R_4^0 \over \sqrt{\a' }}={ R_4^0\over M_P R_5 \sqrt{\a'}}\
\sin(\gamma-\alpha)\ .
\ee
Recalling (\ref{Meff}), i.e. $l_sM_P=2(1+B^2)^{1/2}$, we obtain
\be
g_s^{(B)}=2(1+B^2)^{1/2}\ {R_5 \over R^0_4 \cos(\gamma-\alpha)}\ , \ \ \ \ 
\chi= { R^0_4\over 2(1+B^2)^{1/2}\ R_5 }\ \sin(\gamma-\alpha)\ ,
\label{nni}
\ee
$$ 
\lambda={i\over g_s^{(B)}}+\chi={ R^0_4\over 2(1+B^2)^{1/2}\ R_5 } e^{i(\pi/2+\gamma-\alpha)}\ .
$$
Obtaining a non-zero $\chi $ after the limit $\beta\to \infty $ requires
a rescaling of the radius $R_4^0$ which is different from that
of the $\chi =0$ case considered in \cite {RS}.
So it is convenient to discuss these two cases separately.

\vskip .5cm
\noindent {\bf I)} $g_s\to\infty$ and $\chi=0$

This case can be obtained by setting $\gamma=\alpha$, and ${R^0_4\over
2(1+B^2)^{1/2}}e^{-\beta}\equiv R_4 ={\rm fixed}$. It is worth noting that the compactification radii in the natural OM 
frame \cite{OM} (the coordinate system corresponding to the choice
$g_{\mu\nu}={\rm diag}(-\xi^2,\xi^2,\rho^2,\rho^2,\rho^2,\xi^2)\ $,
for which the natural ``open membrane" metric of the low-energy 
field theory is finite) are both finite, so this is a compactification of OM theory on a finite torus. 
In this way we reproduce results of
\cite{RS}, namely:
\be
G_s=g^{B}_s\sqrt{1-(B^0_{\ 1})^2}\sqrt{1+(B^2_{\ 3})^2}=
2{R_5\over R_4}(1+B^2)=
{\rm finite}\ .
\ee
In the $\beta\to \infty $ limit, the type IIB gauge fields are as follows:
$$
B^0_{\ 1}=1\ ,\ \ \ \ \ B^2_{\ 3}= B\ ,\ \ \ \ \ \chi=0\ ,
$$
\be 
A^0_{\ 1}=0\ ,\ \ \ \ A^2_{\ 3}= {1\over G_s} (1+B^2)\ .
\ee

\vskip .5cm 
\noindent {\bf II)} $g_s\to\infty$ and $\chi={\rm finite}$

{}From eq.~(\ref{nni}) (and recalling that $\alpha\to 0$)
we see that in order to have a finite, non-zero value for $\chi$ 
in the large $\beta $ and large $g_s$ limit, the angle $\gamma $ should approach $ {\pi\over 2}$ and $R_4^0$ should be kept fixed.
This is achieved by scaling these parameters as follows:
\be\label{TTT}
{\pi\over 2}-\gamma=  2\gamma_0\ e^{-\beta}\ ,
\ee
\be
{R^0_4\over M_Pl_s}\equiv R_4={\rm finite}\ ,
\ee
with fixed $\gamma_0$. Then we find
\be
g^{(B)}_s={R_5\over R_4}{e^{\beta}\over B+\gamma_0}\ ,\ \ \ \ \chi={R_4\over R_5}\ ,
\ee
and hence the open string coupling for the non-commutative theory is
\be
G_s={R_5\over R_4}{1+B^2 \over B+\gamma_0}\ .
\ee
The other RR gauge fields (in the $\beta\to \infty$ limit) are
\be
A^0_{\ 1}=A^0_{\ 14}\ (M_P R_4)=-{R_4\over R_5}\tanh\beta \sin\gamma=-\chi \ ,
\ee
$$
A^2_{\ 3}=A^2_{\ 34}\ (M_P R_4)={R_4\over R_5}\sinh\beta \cos\gamma=
{R_4\over R_5}\gamma_0={1+B^2\over G_s}-\chi\ B\ .
$$
This exactly agrees with the asymptotic values of the gauge fields for the
corresponding supergravity configuration for any given $\chi$
(cf. eqs.~(3.2)--(3.4) in \cite{RS}). 
The volume of compactification torus in the natural OM theory frame (see above) 
is proportional to ${\xi\over \rho}\times \cos(\gamma-\alpha)$, which
in the $\beta\to \infty$ limit is  finite.

\newsection{NCOS Theory on a Torus and T-duality}

Non-commutative super Yang-Mills theory on a general $T^2$ torus,
 unlike its commutative counter-part, enjoys the full $SO(2,2;Z)$ T-duality
group of the underlying string theory \cite{SW}. 
In the gauge theory language this is due to the ``Morita equivalence", 
which is an equivalence for the gauge bundles on the non-commutative torus, 
with a proper mapping between the corresponding
gauge groups and couplings, background magnetic fluxes, and  volumes of the two tori
\cite{{CDS},{Connes}}.

Since NCOS theory (with rational $\chi $) is equivalent to NCSYM theory by S-duality,
the T-duality symmetry group of NCOS theory with two spatial dimensions
compactified on a 2-torus must be the same as the T-duality group of NCSYM theory 
$SO(2,2;Z)$. Combined with the S-duality group inherited from type IIB theory, this
gives the U-duality group $SL(3,Z)\times SL(2,Z)$.
In view of the results of section 2,  NCOS theory
on a two-torus can be obtained as a limit of OM theory compactified on a 4-torus $T^4$.
On the face of it, it may seem that the U-duality symmetry group of NCOS theory 
could be one larger than the $SL(3,Z)\times SL(2,Z)$ group. However, this is not the case,
because in order to obtain  NCOS theory in 3+1 dimensions (with no additional Kaluza-Klein states coming  from $d=6$) one must take a zero radius limit of $R_4, R_5$ which  breaks the 
additional  symmetry. Indeed,
after compactifying M-theory on  $T^2\times T^2$ one has
 type IIB string theory in $d=7$, i.e. type IIB on $R^7\times T^3$,
which has a larger U-duality group, $SL(5,Z)$.
To have type IIB string theory in $d=8$, one needs to take an infinite radius limit of one of the $S^1$ in $T^3$
(which, in M-theory variables, corresponds to the zero area limit of one of the 2-torus). This reduces the U-duality group to $SL(3,Z)\times SL(2,Z)$.

Consider NCSYM theory on a 2-torus, with $x_2,x_3$  periodic coordinates, $x_2=x_2+2\pi R_2$, $x_3=x_3+2\pi R_3$.
The torus volume is $V=4\pi^2 R_2 R_3$.
When the dimensionless non-commutative parameter $\Theta=\theta^{23} / V$
is a rational number, the theory is equivalent to an ordinary
SYM in the presence of a magnetic flux \cite{CDS}.
This equivalence can be understood from the T-duality symmetry
of type II string theory.

Because the NCOS theories too arise as a limit of type  IIB string theory in some
background, there  will be T-duality connections when they are compactified on a two-torus.
 Type IIB string theory has in addition  
the strong-weak  $SL(2,Z)$ duality symmetry. 
In ref.~\cite{RS} we described how the S-duality $SL(2,Z)$ transformations act on
the parameters  of the (3+1) dimensional NCOS theories. 
The full U-duality group of type IIB superstring on a 2-torus is
$SL(3,Z)\times SL(2,Z)$. 
General type IIB $SL(3)\times SL(2)$ transformations
will induce some transformations between different NCOS theories,
i.e. it will map one into another with transformed parameters.

Let us now investigate the implications of T-duality in more detail. 
NCOS theories on a rectangular two-torus are specified by parameters:
$$
(\a',\ G_s,\ \theta^{23},\ \chi \ ;\ N,\ R_1,\ R_2)
$$
where $\chi $ arises from the expectation value of the RR scalar, giving rise to
a $\theta_{\rm QCD}$ term in the low energy effective lagrangian.
If $\chi $ is rational, it was shown that the theory is equivalent
(by an $SL(2,Z)$ S-duality transformation) to NCSYM theory \cite{RS}.
In particular, if $\chi=0$, the NCOS theory with $\theta^{23}=0$
   is S-dual to 
NCSYM theory by the simple S-duality transformation $\lambda\to -1/\lambda $ that inverts the gauge coupling \cite{GMMS}. 
Thus NCOS theory 
with $\chi=\theta^{23}=0$ can be viewed as the strong coupling limit
 of NCSYM theory.
For the NCOS theory on the torus, 
the non-commutativity in the $x^0$-$x^1$ 
directions  is characterized by the parameter $\theta^{01}=2\pi \a' =V\Theta $.
Although being a perturbative symmetry, T-duality is believed to hold even for 
strong coupling. This indicates that  NCOS theory with rational $2\pi\a'/V$
and $\chi=0=\theta^{23}$
must be equivalent to an ordinary Yang-Mills theory
with a flux. 

The precise background fields can be found by transforming the supergravity dual background.
One can start with the supergravity background describing NCSYM theory \cite{hashi,MR}
and apply
a suitable T-duality transformation leading to a geometry $AdS_5\times S^5$ with constant dilaton and constant $B_{23}$ field \cite{hasss}. 
By S-duality, this can be converted into $AdS_5\times S^5$ with constant dilaton and 
constant $A_{23}$ field.
 Alternatively, one can start with the supergravity dual background describing
NCOS theory \cite{MR} and look for an $SL(3,Z)$ transformation that
leads to $AdS_5\times S^5$ with constant dilaton.
The S-duality and T-duality $SL(2,Z)$ matrices can be embedded into $SL(3,Z)$ matrices as follows:
$$
g_{S}=\left(\matrix{a& b& 0 \cr c& d& 0\cr 0& 0& 1}\right)\ ,
\ \ \ \ \ g_T=\left(\matrix{1& 0& 0 \cr 0& a& b\cr 0& c& d}\right)\ .
$$
The transformation that takes NCOS theory with rational $\Theta $ 
into ordinary YM theory is of the form
\be
g_{U}=g_S\ g_T\ g_S= \left(\matrix{a& 0& b \cr 0& 1& 0\cr c& 0& d}\right)\ ,
\label{aaaaa}
\ee
where $g_S$ is as above with $a=d=0$, $c=-b=1$. 
Then, the transformed dilaton is constant provided
$$
a+bA_{23}^\infty =0\ ,\ \ 
$$
where $A_{23}$ is the non-vanishing component of the RR two form.
This implies a rational asymptotic value for the $A_{23}$ field, 
$A_{23}^\infty=-{b\over a}$.
We omit the details of the calculation
(the transformation properties of type IIB supergravity fields under $SL(3)$ can be found
in \cite{KP}). Thus the required transformation  is
an element of $SL(3,Z)$ which is not in $SL(2,Z)_T$ or $SL(2,Z)_S$ subgroups.
This is already clear from eq.~(\ref{aaaaa}).

Let us now consider T-duality transformations on NCOS theories with 
$\theta^{23}\neq 0$ (an independent discussion has appeared in \cite{gremm} while
this paper was in preparation).
It should be noted that, for irrational $\chi $, NCOS theories are not  related to a NCSYM theory, so to study  T-duality of such NCOS theories  is convenient to start with the
appropriate type IIB configuration and then take a limit leading to a NCOS theory. 
Once the moduli parameters of the compactified theory are specified,
it is easy to obtain the  T-duality transformation properties.
For the case of our interest, (3+1) dimensional NCOS with parameters, 
$(\a',\ G_s,\ \theta ^{23} ,\ \chi)$, we consider the
compactification of the $\theta$-plane $x_2$-$x_3$ on a non-commutative torus 
two torus  $T^2_{\theta}$.
This theory can be realized as some particular limit of type IIB string theory
in the presence of a $(D3,(F,D1))$ -brane
system compactified on the two torus. 
We denote the complex and Kahler parameters of that torus by $\tau$ and
$\rho$, respectively. Then the brane bound state
is characterized by two integers $m,\ N$, whose ratio is proportional to the RR charge density corresponding to
D-strings \cite{{AAS1},{AAS2}}. 

We choose coordinates so that the components of the 
closed string metric parallel to the brane bound state are 
$(-{1\over 1-E^2},\ {1\over 1-E^2},\ 1,\ 1)$. 
Along the lines of  \cite{{AAS2},{Dip}}, the spectrum of the open strings attached to the brane bound state is 
\be
\a' M^2={|r+q\tau|^2\over \tau_2}\ {\rho_2\over |m+N\rho|^2}+ {\rm Oscil.}\ ,
\ee 
where $\tau_2$ and $\rho_2$ are the imaginary parts of $\tau$ and $\rho$, $r$ and $q$ are two integer parameters
representing the winding and momentum modes of open strings, respectively. We see  that the 
zero mode part of the spectrum is manifiestly invariant
under the T-duality group $SO(2,2;Z)\sim SL(2,Z)_{\tau}\times\ SL(2,Z)_{\rho}$. 
The other open string parameters, i.e. $\theta^{\mu\nu}$ and $G_s$ are \cite{RS}
\be
\theta^{01}=-\theta^{10}=2\pi\a' E\ \ \ \ \ \    
\theta^{23}=-\theta^{32}=2\pi\a'{B\over 1+B^2}\ ,
\ee
\be
G_s=g_s\sqrt{(1-E^2)(1+B^2)}\ .    
\ee
Let us now take the $E\to 1$ limit while keeping $\a'$, $G_s$ and the volume of the torus fixed. This  leads
to a NCOS theory on $T^2_{\theta}$ defined by parameters:
$(\a',\ G_s,\ \theta ^{23},\ \chi; m ,\ N,\ R_1,\ R_2)$.
The $SL(2,Z)_{\tau}$ part consists of transformations under which
$\theta^{23},\ m,\ N $, $\chi$  and $ G_s$ are invariant; it only acts on
the torus metric (and $r$ and $q$ modes). 
Other  transformations are generated by $\left(\matrix{a & b \cr c & d} 
\right)\in SL(2,Z)_\rho $, which  act on the torus volume $V,\ \theta,\ G_s$
and $(m,\ N)$
as \cite{SW}
$$
V'=V\ (a+b\Theta)^2\ \ \ , G'_s=G_s\ (a+b\Theta)\ , \ \ \ \
\Theta={\theta^{23}\over V}\ ,
$$
\be
\Theta'={c+d\Theta\over a+b\Theta}\ ,
\ee
$$
\left(\matrix {m'\cr N'}\right)=\left(\matrix{a& b\cr c&
d}\right)\left(\matrix{m \cr N}\right)\ .
$$
Thus, under this transformation, a NCOS theory is mapped into another 
NCOS theory with the same $\a'$ and $\chi$
parameters, while all other moduli are transformed as above.
For the special case of rational $\Theta$ there is a T-duality
under which $\Theta$ vanishes and the resulting theory is NCOS theory with $\theta^{23}=0$.

\newsection{ Solitons of NCSYM theories from D-branes }

Soliton solutions of NCSYM theories can be
described in terms of brane configurations in string theory. In the usual (commutative) case, the (3+1) SYM theory
can be realized as the low-energy worldvolume theory of a D3-brane 
(we assume the D3 brane lies along $0123$ directions). From this point of
view, the BPS monopole solution 
is described by a D-string ending on the D3-brane, e.g. 
a D-string along the $04$ direction. 
This picture can be generalized to obtain a description of monopoles in non-commutative 
super Yang-Mills theories by considering a D3-brane along
$0123$ directions with a $B_{23}$ field turned on. 
The worldvolume theory of this brane is then NCSYM with
non-commutative parameter $\theta_{23}$. 
As noted in \cite{Hashi, GN} the requirement of unbroken supersymmetry
determines the incident angle of the D1 string on the D3 brane.
A part of the present discussion has already appeared in \cite{Hashi}.
In addition, we give an explicit solution for the dyonic case, and discuss the M-theory 
origin of monopole and dyon configurations. 

\subsection{Monopoles and dyons in non-commutative theories}

Let $Q_L$ and $Q_R$ be the 32 conserved supercharges of type IIB theory. 
The D3-brane with a $B_{23}$ field 
preserves 16 supercharges, given by the supercurrent 
$$
\epsilon_L\ Q_L +\epsilon_R\ Q_R\ , 
$$
where $\epsilon_L$ and $\epsilon_R$ are 16 component Killing spinors 
which satisfy \cite{config}
\be\label{SUSY1}
\Gamma^{0123}\big({1\over \sqrt{1+B^2}}-{B\over \sqrt{1+B^2}}\Gamma^{23}\bigr)\epsilon_L=\epsilon_R\ .
\ee
In the above equation $\Gamma$ matrices are the 10 dimensional Dirac matrices\footnote{We
choose conventions in which the closed string metric is $\eta_{\mu\nu}$, with signature $(-,+,+,...)$.}:
$$
\{ \Gamma^{\mu},\ \Gamma^{\nu}\} =2\eta^{\mu\nu }\ .
$$
Let us now add the D-string to this brane system. Suppose that the D-string is along direction $4'$, which makes 
an angle $\phi_0$ with respect to the $X^4$ direction in the $14$ plane, then the unbroken supersymmetry generators are given by Majorana-Weyl spinors satisfying
\be\label{SUSY2}
\Gamma^{04}e^{{\phi_0}\Gamma^{14}}\epsilon_L=\Gamma^{04}\bigl({\cos\phi_0}-{\sin\phi_0}\Gamma^{14}\bigr)\epsilon_L=
\epsilon_R\ .
\ee
In order that eqs.(\ref{SUSY1}) and (\ref{SUSY2}) have simultaneous solutions, i.e. the BPS condition,
we need that 
\be\label{SUSY4}
\Gamma^{1234}\bigl({\cos\phi_0}+{\sin\phi_0}\Gamma^{14}\bigr)
\big({1\over \sqrt{1+B^2}}-{B\over \sqrt{1+B^2}}\Gamma^{23}\bigr)\epsilon_L=-\epsilon_L\ .
\ee
Since $\Gamma^{14}$ and $\Gamma^{23}$ are commuting, in a proper representation they can be chosen as
\begin{eqnarray}
\Gamma^{12} & = & i \: \hbox {diag} (+ {\bf 1}_8, -{\bf 1}_8),
\nonumber \\
\Gamma^{34} & = & i \: \hbox {diag} (+ {\bf 1}_4, -{\bf 1}_4, +{\bf 1}_4, -{\bf 1}_4)\ ,
\label{repre}
\end{eqnarray}
where by ${\bf 1}_D$ we denote $D$ dimensional identity matrices. 
Then eq.~(\ref{SUSY4}) has solutions provided:
\be\label{tilte}
\tan\phi_0=\pm B\ .
\ee
Therefore the D-string should be tilted with respect to D3-brane \cite{Hashi}. 
The $+$ and $-$ signs correspond to monopole
and anti-monopole respectively. This is indeed the same condition
found by field theory arguments \cite{GN}. 
Let us now find the monopole mass, which is related to the string tension. 
If we denote the tension of open F-strings ending on the NC D3-brane
by ${1\over \alpha'}$, then, according to the picture of \cite{GN}, the 
``shadow" or projection of a D-string on the D3-brane worldvolume carries an  energy per unit
length $t_{\rm eff}$ given by
\be\label{tens}
t_{\rm eff}={T_{{\rm D-string}}\over \sin\phi_0} ={1 \over \alpha'g_s}\ {1\over \sin\phi_0}\ .
\ee
The appearance of the sine factor is a consequence of the D-string tilt. 
Using eq.(\ref{tilte}) we have
\be
t_{\rm eff}={B \over \alpha'g_s\sqrt{1+B^2}}={1+B^2\over \alpha' B}\ {1\over g_s\sqrt{1+B^2}}={1\over g_{YM}^2\theta}\ ,
\ee
which is in exact agreement with the results of \cite{GN}.

One can similarly describe NCSYM $(p,q)$ dyons (with (0,1) corresponding to
a D-string). Along the arguments 
of \cite{{Bound},{AS}},
the $(p,q)$-string can be realized in string theory as a D-string with a non-zero electric background, $E$, where 
\be\label{pq} 
{E \over \sqrt{1-E^2}}={p\over q}\ g_s\ . 
\ee 
Then, the supersymmetry preserved by a $(p,q)$-string
along the $4'$ direction, instead of eq.~(\ref{SUSY2}), is given by 
\be\label{SUSY3}
\Gamma^{04'}\big({1\over \sqrt{1-E^2}}-{E\over \sqrt{1-E^2}}\Gamma^{04'}\bigr)\epsilon_L=\epsilon_R\ , 
\ee 
where
\be 
\Gamma^{4'}=\Gamma^{4}\big({\cos\phi}-{\sin\phi}\Gamma^{14}\bigr)\ . 
\ee 

Eqs.(\ref{SUSY1}) and (\ref{SUSY3}) will have simultaneous solutions provided the matrix $A$:
\be
A=\Gamma^{1234}({1\over \sqrt{1-E^2}}-{E\over \sqrt{1-E^2}}\Gamma^{04'}\bigr)
\bigl({\cos\phi}+{\sin\phi}\Gamma^{14}\bigr)
\big({1\over \sqrt{1+B^2}}-{B\over \sqrt{1+B^2}}\Gamma^{23}\bigr)\ ,
\ee
has some eigenvalues equal to $-1$. By straightforward matrix algebra we find that 
this condition is satisfied only when
\be\label{Phi}
\sin\phi= {B\over \sqrt{1+B^2}}\ \sqrt{1-E^2}\ ,
\ee
where $E$ is related to $(p,q)$ as in eq.~(\ref{pq}). 
We see that the incident angle for a $(p,q)$ string is in general {\it less} than 
that of a D-string. The angle $\phi$
does not depend on the sign of $p$, and only the combination $({p\over q})^2$ appears in (\ref{Phi}). In the special cases $p=0$ we recover the monopole case considered
above. The case $B=0$ corresponds to $(p,q)$-dyons of a commutative SYM theory. 

The effective tension of a $(p,q)$-dyonic state
of NCSYM theory will be given by
\be\label{Tpq}
t_{(p,q)}={T_{(p,q)}\over \sin\phi}={q\over g^2_{YM}\theta}\ \bigl(1+({p\over q})^2g_s^2\bigr)\ ,
\ee
where the closed string coupling $g_s$ is related to $g_{YM}$ and $\theta $ 
by $g_{YM}=g_s\sqrt{1+B^2}$ and
$\theta={2\pi\alpha' B\over 1+B^2}$.


The present treatment makes use of the perturbative ($g_s \ll 1$)
description of D-branes in terms of Dirichlet (and Neumann) 
boundary conditions on open strings. 
The brane intersections discussed here are BPS, so they 
exist also in the regime $g_s\gg 1 $.
In the strong coupling limit, one can try to use the S-dual picture, corresponding
to NCOS theory with parameter $\theta^{01}$. However, it should be noted that
the usual noncommutative open strings are not the S-duals of the BPS configurations we
discussed above (in particular, they are electrically neutral, being dipoles having two ends with opposite  charges \cite{Dip}). They are non-BPS configurations whose
tensions are twice more than the BPS configuration that is S-dual to monopoles.

For the special case of $4+1$ dimensional NCSYM theory, which in the strong coupling limit
is OM theory, the $(p,q)$ dyon state corresponds to a (D2-brane, F-string) bound state intersecting a D4-brane.
In OM theory, the $(p,q)$ dyon state corresponds to open membrane states ending on the critical M2-M5 brane bound state.

\subsection{ Description of monopoles and dyons from six dimensions}

To describe monopole states of $(3+1)$ dimensional NCSYM theory from 
OM theory, we start with an open membrane ending on M5-brane
in the presence of a $H$ field, make dimensional reduction and T-duality.
We will follow the conventions of section 2.

The discussion of the previous section regarding the supersymmetry preserved by 
the D3-D1 bound state can be easily generalized to this case
of an M5-brane with a background $H$-field.
This system can be viewed as a non-marginal bound state of (M2,M5)-branes 
\cite{rutse,SorT}, and in this section we use their conventions for $\Gamma$
matrices. 
A similar discussion is given in \cite{OhT}. 
Let $\epsilon$ be the asymptotic value of the 32 component Killing
spinor of an eleven-dimensional supergravity solution corresponding to a (M5-M2) bound state, describing a M5-brane
along $012345$ directions with a non-zero $H_{234}$ field. We have
$$
\Gamma^{012345}({1\over \sqrt{1+H^2}}-\Gamma^{234}{H\over \sqrt{1+H^2}})\epsilon=\epsilon\ ,
$$
where $H\equiv H_{234}$. The above equation can also be
expressed in terms of $H_0\equiv H_{015}$,
$$
\Gamma^{012345}({1\over \sqrt{1-H_0^2}}+\Gamma^{015}{H_0\over \sqrt{1-H_0^2}})\epsilon=\epsilon\ ,
$$
provided that $H_0={H\over \sqrt{1+H^2}}$, which is nothing but the {\it
self-duality} condition. The space of solutions of these equations is 16
dimensional, i.e. the configuration preserves ${1\over 2}$ of the original
supersymmetries.
Now we add an intersecting open membrane to the above (M5-M2) bound state. 
Assuming this membrane lies on 
the $046'$ directions, with
$6'$ being a direction on the $16$ plane making an angle $\phi_0$ with respect to $6$,
the Killing spinor should satisfy
$$
\Gamma^{046}(\cos\phi_0-\Gamma^{16}\sin\phi_0)\epsilon=\epsilon\ .
$$
The condition for having a BPS brane intersection tells us that 
$$
\tan\phi_0=\pm H\ ,
$$ 
which in terms of the $\beta$ parameter of OM theory is 
\be
\tan\phi_0=\pm\sinh\beta \ .
\ee
One can also consider a more general case in which the open membrane is along $04'6'$. 
In this case the supersymmetry conditions lead to (\ref{phipsi}).

To obtain a D-string (monopole) intersecting a D3-brane in the presence of a $B_{23}$ field, 
we make dimensional reduction and T-duality as in section 2, setting 
both $\alpha$ and $\gamma$ to zero.
Reducing along $x^5$, we will find a type IIA configuration of D4-branes along 
$01234$ with a non-zero $B_{23}$ field. 
The open membrane becomes a D2-brane along $046'$. 
By T-duality along $x^4$,
we end up with a type IIB D3-brane along $0123$, with a non-zero $B_{23}$ and a D-string along
$06'$, which is precisely the monopole state of the previous section.

The dyonic $(p,q)$ states of a $(3+1)$ NCSYM theory can also be realized from OM theory. 
Consider an M2-brane lying along $04'6'$, with $4'$ and $6'$ being 
arbitrary directions in $45$ and $16$ planes making angles $\psi$ and $\phi$ with 
respect to $4$ and $6$ directions,
respectively. Reducing this membrane along the $5$ direction
leads to a (D2-brane,F-string) bound state of type IIA string theory, 
lying on the $046'$ direction (with the F-strings
oriented along the $06'$ directions). The dyonic $(p,q)$ state 
of $(3+1)$ NCSYM theory is finally obtained upon
T-duality along $x^4$ direction. In this way one finds that $p\over q$ 
is related to $\psi$ as
\be
\tan\psi={p\over q}g_s\ .
\ee
Summarizing, the dyon state of NCSYM theory can be described in eleven dimensions by an M5-brane with a non-zero $H_{234}$ field component (which by self-duality requires as well a $H_{015}$ 
component) with an open membrane along $04'6'$ directions ending on the M5 brane.
For a given $\psi$, the supersymmetry conditions fix the angle $\phi$ to be given by
\be\label{phipsi}
\sin\phi=\tanh\beta\cos\psi\ .
\ee

\subsection{ Supergravity solution}
\def\del{\partial }

Let us first recall the supergravity solution that is dual to large
$N$ NCOS theories \cite{MR}.
We follow the notation of eq.~(5.2) of \cite{RS}.
\be
ds^2=\a' f^{1/2}\bigg[{u^4\over R^4}(-dx_0^2+dx_1^2)+
{\hat h}{u^4\over R^4}(dx_2^2+dx_3^2)+du^2+u^2d\Omega^2_5\bigg]\ ,
\label{ppu}
\ee
$$
B_{01}=-\a' {u^4\over R^4}\ \ ,\ \ \ \ \ 
B_{23}=\a' {u^4\over R^4}{\hat h}\tan\a\ .
$$
In ref. \cite{SW} it was shown that open strings in the presence of 
a constant  gauge field $B$ and a {\it flat}  metric $g_{\mu\nu}$  can be effectively  described as open strings in a background with  $B=0$, but with a metric $G_{\mu\nu}$ and non-commuting coordinates with
 $[x^\mu ,x^\nu]=i\theta^{\mu\nu}$, related to $g_{\mu\nu}$ and $B$ by 
$$
(G+\theta)^{\mu\nu}=((g+B)^{-1})^{\mu\nu}\ .
$$
In type IIB superstring theory, the presence of  D3 branes and gauge fields are 
 sources for gravitational
fields, which curve  the metric and modify other background fields,
e.g. as in (\ref{ppu}). It is interesting to extend the above definition of $G_{\mu\nu}$ 
and $\theta^{\mu\nu}$ --which are natural objects from the
point of view of the dual non-commutative theory-- to this case. We find
\be
G^{\mu\nu}=\a{'^{-1}}f^{1/2}\ \eta^{\mu\nu}\ ,
\ee
\be
\theta^{01}=\a'\ \ ,\ \ \ \ \theta^{23}=\a'\tan\a \ .
\ee
Interestingly, although the $B$ field is $u$-dependent, $\theta ^{\mu\nu}$ is constant.
Unlike $g_{\mu\nu}$, the open string metric is conformal to the $d=4$ Minkowski metric.

\medskip

A supergravity background representing a D-string intersecting
a D3 brane in the presence of a $B$-field can be constructed as follows.
For $B=0$, one has the 1/4 supersymmetric solution representing
orthogonal intersection of D-string and D3 brane, given by
\def\ha{ {1\over 2} }
$$
ds^2=f_1^{-\ha}\bigg( f_3^{-\ha }\big[-dx_0^2+f_1(dx_1^2+d x_2^2+dx_3^2)\big]
+f_3^{\ha } \big[ dx_4^2 +f_1(dr^2+r^2 d\Omega_4^2)\big]\bigg)\ ,
$$
\be
e^{2\phi }=g^2 f_1\ ,\ \ \ f_{1,3}=1+ {Q_{1,3}\over r^3}\ ,
\ee
$$
F_{04r}={1\over g}\ \del_r f_1^{-1} \ ,\ \ \ \ 
F_{0123r}={1\over g} \ \del_r f_3^{-1}
$$
Now we make a T-duality transformation in the $x_2$ direction, a rotation in the
plane $(x_2,x_3)$, and T-duality in the new $x_2$ coordinate. We obtain
$$
ds^2=f_1^{-\ha}\bigg( f_3^{-\ha }\big[-dx_0^2+f_1
dx_1^2+f_1h (d x_2^2+dx_3^2 ) \big]
+f_3^{\ha } \big[ dx_4^2 +f_1(dr^2+r^2 d\Omega_4^2)\big]\bigg)\ ,
$$
\be
e^{2\phi }=g^2 f_1 h \ ,\ \ \ h^{-1}=\sin^2\theta \ {f_1\over f_3}
+\cos^2\theta\ ,
\ee
$$
B_{23}={\sin\theta\over \cos\theta }\ {f_1\over f_3}h\ ,
$$
$$
F_{04r}= {1\over g}\ \cos\theta \del_r f_1^{-1} \ ,\ \ \ \ 
F_{0123r}={1\over g}\ \cos\theta \ h \ \del_r f_3^{-1}\ ,\ \ \ \ F_{0234r}={1\over g} \ \sin\theta\ \ h\ \del_r f_3^{-1}\ .
$$
The charge density of the string is distributed along the axes $x_4$~;
the charge density of the D3 brane is distributed in the volume
$(\tilde x_1,x_2,x_3)$, where 
$$
\tilde x_1=x_1\cos\theta +x_4\sin\theta\ .
$$
Thus the D-string is at an angle ${\pi\over 2}- \theta $ with respect to the
D3 brane. The asymptotic value of the $B$-field
is $B_{23}^\infty=\tan\theta $. This is consistent with the
field theory analysis, and with the discussion of supersymmetry of section 4.1.

Let us now consider the decoupling limit. 
Let $x_4$ be a periodic coordinate $x_4=L\theta_4 $,  with $\theta_4 =\theta_4+2\pi $,
and write
$$
f_1=1+{Q_1\over r^3}\ ,\ \ \ \ f_3=1+ { {\a' }^2 R^4\over 2\pi L r^3}
$$
Now we scale variables as follows
$$
r=\a' u\ ,\ \ \ \ L=\a' \tilde L\ ,\ \ \ \tan\theta ={\tilde b\over\a' }\ ,
$$
$$
x_{2,3}=\a' \tilde x_{2,3}\ ,\ \ \ 
Q_1={\a'}^3 \tilde Q_1\ ,\ \ \ Q_3={ {\a' }^2 R^4\over 2\pi L }=\a' R^4_0\ ,
$$
and take the limit $\a' \to 0$ with fixed $u,\tilde b , R_0,\tilde L,\tilde x_{2,3}$.
The resulting metric has following the form
$$
ds^2=\a ' f_1^{1\over 2}
\bigg[ {u^{ 3\over 2 }\over R_0^2}\big[ -f_1^{-1 } dx_0^2
+dx_1^2 +\hat h(dx_2^2+dx_3^2)\big]\ ,
$$
$$
+\ {R_0^2\tilde L^2\over f_1 u^{3\over 2}}
d\theta_4^2 + {R_0^2\over u^{3\over 2}}(du^2+u^2d\Omega_4^2)\bigg]
$$
with
$$
\hat h^{-1}=1+a^3 u^3\ ,\ \ \ \ a^3={\tilde b^2\over \tilde b^2\tilde Q_1+R_0^4}\ .
$$

\subsection{Instanton solution}

Finally, we point out another solution in Euclidean space
which may be relevant to the study of instantons in non-commutative
super Yang-Mills theories \cite{NS}.
Consider the Euclidean
solution representing a bound state of D(-1) and D3 brane.
It is given by \cite{liut}
$$
ds^2=H^{1/2}\bigg[ f^{-1/2}\big[ dx_0^2+dx_1^2+dx_2^2+dx_3^2\big]+
f^{1/2}\big( dr^2+r^2d\Omega_5^2\big)\bigg]\ ,
$$
\be
e^{2\phi}=g^2 H^2\ ,\ \ \ \ \ \chi={i \over g} H^{-1}\ \ ,
\ee
$$
F_{0123r}={i\over g} \del f^{-1}\ ,
$$
$$
f=1+{{\a'} ^2R^4\over r^4}\ ,\ \ \ \ \ H=1+{r_0^4\over r^4}\ .
$$
A $B$ field can be introduced as in the previous subsection. We perform T-duality in
the direction $x_2$, make a rotation in the plane 2-3, and then
T-duality in $x_3$. Using the standard T-duality rules,
we find the following solution
$$
ds^2=H^{1/2}\bigg[ f^{-1/2}\big[ dx_0^2+dx_1^2+h(dx_2^2+dx_3^2)\big]+
f^{1/2}\big( dr^2+r^2d\Omega_5^2\big)\bigg]\ ,
$$
$$
e^{2\phi}=g^2 \ h \ H^2\ ,\ \ \ \ \ \chi={i\over g} \cos\theta\ H^{-1}\ ,
$$
\be
B_{23}=\tan\theta \ {h\over f}\ H\ ,\ \ \ \ A_{23}=i \sin\theta\ H^{-1}\ ,
\label{rrre}
\ee
$$
F_{0123r}={i\over g} \cos\theta \ h \del f^{-1}\ ,
\ \ \ \ h={f\over H \sin^2\theta + f\cos^2\theta } \ .
$$
The decoupling limit of this solution is obtained by
dropping the ``1" in the function
$f$, but not in $H$
(dropping the ``1" in both functions leads to constant $h$ and $B$,
that is, a solution which is essentially equivalent to the $B=0$ case).

It is interesting to note that there is a similar solution with
real gauge fields, but for a spacetime signature $(--++)$, which
exists only in the non-commutative theory (i.e. only for $B\neq 0$).
This can be obtained as follows.
First, we make the shift $\theta = \alpha +\pi/2 $
so that $\sin\theta \to \cos\alpha $, $\cos\theta\to -\sin\alpha $.
Then we change
$$
\alpha\to i\alpha\ ,\ \ \ \ x_0\to ix_0\ ,\ \ \ \ x_2\to ix_2\ .
$$
The resulting solution is
$$
ds^2=H^{1/2}\bigg[ f^{-1/2}\big[ -dx_0^2+dx_1^2+h(-dx_2^2+dx_3^2)\big]+
f^{1/2}\big( dr^2+r^2d\Omega_5^2\big)\bigg]\ ,
$$
$$
e^{2\phi}=g^2 \ h \ H^2\ ,\ \ \ \ \ \chi={1\over g} \sinh\a\ H^{-1}\ ,
$$
\be
B_{23}=\coth\a \ {h\over f}\ H\ ,\ \ \ \ A_{23}= \cosh\a \ H^{-1}\ ,
\label{rre}
\ee
$$
F_{0123r}={1\over g} \sinh\a \ h \del f^{-1}\ ,
\ \ \ \ h={f\over H \cosh^2\a - f\sinh^2\a } \ .
$$
In this solution all fields are real. 
In the commutative case $B_{23}=0$, corresponding to $\theta =0$ in (\ref{rrre}), one has
$\chi ={i\over g} H$, and it is not possible to make it real.
This can also be seen from the fact that 
$B_{23}=\coth\a {hH\over f} \neq 0$ for all $\a $, so  
the above solution (\ref{rre}) exists only for $B_{23}\neq 0$.

\newsection{Potential between monopole and antimonopole}

Consider a D3 brane in presence of a magnetic field $B_{23}$.
The ends of a D-string attached to the D3 brane 
represent a monopole $m$ and an antimonopole $\bar m$
from the standpoint of the
3+1 dimensional non-commutative field theory. The potential between 
$m$ and $\bar m$ can be obtained in the large $N$ limit by using the 
dual supergravity description and computing a Wilson loop.
The supergravity background we use is given by eq.~(2.7) in \cite{MR}.
We consider a D-string configuration in this geometry
of the form $x^0=\tau, x^2=\sigma $, i.e.  
 a string lying on a plane  orthogonal to
the magnetic field.
The Born-Infeld action for the D-string in this background is
\be 
S={T\over 2\pi \hat g }\int d\s \sqrt{(1+a^4u^4) 
(\partial_\s u)^2 +{u^4\over R^4} }\ ,\ \ 
\ee
This action formally coincides with the case $\alpha =0$ of
eq.~(5.4) in \cite{RS}, representing the Nambu-Goto action
for a fundamental string attached to a D3 brane in the presence of a $B_{01}$ field
(this was used to obtain  a quark-antiquark potential in NCOS theory).
The present $m-\bar m$ case can be regarded as the S-dual situation.
The solution to the equations of motion is
\be
(\partial_\s u)=
{u^2 \over R^2 \sqrt{1+a^4 u^4}} \sqrt{ {u^4\over u^4_0}-1}\ ,
\ee
where $u_0$ represents the minimum value of $u$ reached by the string,
which has its ends at $u= \infty $. 
It is related to the 
distance $L$ between $m$ and $\bar m$ by the formula:
\be
L(u_0)=\int dx_2= {2 R^2 \over u_0}
\int_1^{\infty}{dy \over y^2}{ \sqrt{1+a^4u_0^4 y^4}\over
\sqrt{y^4-1 } }\ .
\label{lle}
\ee
This equation defines $u_0=u_0(L)$.
The monopole-antimonopole potential $V(L)$ is obtained from
$$
V(L)=V[u_0(L)]={1\over T}\big( S(u_0)- S(0)\big)\ ,
$$
i.e.
$$
V(L)={u_0 \over \pi \hat g}
\int_1^{\infty} dy
\sqrt{1+u_0^4a^4y^4}\left( 
{y^2 \over \sqrt{y^4-1 } } -1\right)
- {u_0 \over \pi \hat g} \int_0^1 dy \sqrt{1+a^4u_0^4y^4}\ .
$$
Being formally the same as the quark-antiquark potential
of the S-dual electric non commutative theory,
it will be given by the same formulas as in \cite{RS},
with the appropriate correction in the $R$ (coupling) dependence:
\be
u_0={a_0R^2\over L}+ a_1 a_0^3 {R^8\over L^4} +O(1/L^5)\ ,
\ \ \ \ \ R^4=4\pi \hat g N=  g_{YM}^2 N \equiv \lambda \ ,
\label{zxx}
\ee
$$
a_0=2\int_1^\infty {dy\over y^2}{1\over \sqrt{y^4-1}
}={2\sqrt{2}\pi^{3/2}\over
\Gamma({1\over 4})^2 }\ ,\ \ \ \ \ a_1={\Gamma({1\over 4})^2\over
3\sqrt{\pi } }\ ,
$$
\be
V(L)=-{4\pi N \over \lambda }\bigg( {c_0  \sqrt{\lambda }\over L} - 
{c_1 \lambda^2 \over L^4}\bigg) +O(1/L^5)\ ,
\label{vvll}
\ee
$$
c_0={4\pi ^2\over \Gamma({1\over 4})^4 }\ ,\ \ \ \ 
c_1={16 \pi^{9/2}\over 3 \Gamma({1\over 4})^6 }\ .
$$
The leading term in the expansion in powers of $1/L$ 
coincides with the
commutative case \cite{minahan}. The subleading corrections
are due to non-commutativity.
The potential $V(L)$ can be computed numerically for all $L$.
{}From the plot obtained in \cite{RS} for the dual quark-antiquark configuration,
one can see 
that the curve $V(L)$ terminates at a certain 
$L=L_{\rm min}$. This may indicate that
 for $L<L_{\rm min}$ the potential becomes constant,
since the only solution is that  of two separate D-strings.

\newsection{Conclusions}

In this paper we have studied some non-perturbative aspects of NCOS
theories and their OM theory origin. 
In particular, we have obtained  monopole and dyonic 
solutions of (3+1) dimensional NCOS from the 
intersecting brane picture and found their mass density.
Their stability is supported by the unbroken supersymmetries determined in sect.~4.
In addition we have also presented a supergravity solution related to these
 non-perturbative states, and a solution representing instantons in NCSYM theory.

Compactifying OM theory on a general torus, we have shown that the full four parameter 
class of (3+1) dimensional NCOS theories
can be obtained from toroidal compactification of OM theory. 
We presented some evidence that there is a T-duality $SO(2,2;Z)$ group
acting on  NCOS theories compactified on a general non-commutative two torus, 
although being an {\it open string} theory without any closed string sector.
Altogether, the full set of transformations
acting on  NCOS theories on $T^2$ are induced by the $SL(3,Z)\times SL(2,Z)$ 
U-duality group of type IIB superstring theory. 
A specific $SL(3)$ transformation maps NCOS theory with rational $2\pi\a'/V$
and $\chi=0=\theta^{23}$
into an ordinary Yang-Mills theory
with a magnetic flux.

\bigskip\bigskip\bigskip

{\bf Acknowledgements}

J.R. would like to thank CERN for hospitality
during the course of this work.
The work of M.M. Sh-J. was partly supported by the EC contract
no. ERBFMRX-CT 96-0090.

\end{document}